# Phase diagram of the $NdFe_{1-x}Rh_xAsO$ superconductor

H. Y. Shi, X. L. Wang, T.-L. Xia, Q. M. Zhang, X. Q. Wang, T.-S. Zhao*

Department of Physics, Renmin University of China, Beijing 100872, P. R. China

Polycrystalline samples with a nominal composition of $NdFe_{1-x}Rh_xAsO$ ($0 \leq x \leq 0.25$) were synthesized using a solid state reaction method. Bulk superconductivity with a maximum $T_C = 13$ K is observed in the $x = 0.1$ sample. A temperature-composition phase diagram is established for the $NdFe_{1-x}Rh_xAsO$ system based on the electrical resistivity and magnetization measurements. A first-order-like transition from an antiferromagnetic state to the superconducting state at a critical Rh-doping point $x_C \approx 0.045$ is observed in the present phase diagram. The value for the upper critical field $\mu_0 H_{c2}(0)$ is estimated to be about 26 T for the $x = 0.10$ sample by using the Werthamer–Helfand–Hohenberg theory.

*Corresponding author. E-mail: tszhao@ruc.edu.cn

Since the recent discovery of superconductivity in LaFeAs(O, F) with transition temperature up to $T_C$ = 26 K [1], the FeAs-based superconductors have attracted much attention. The superconducting (SC) transition temperature $T_C$ was quickly increased up to as high as 55 K in *R*FeAs(O, F) (*R* = Pr, Nd, Ce, Sm, Gd ) by replacing La with other rare earth with smaller ionic radius [2-5]. Moreover, a 3*d* metal Co or Ni substitution for Fe in *R*FeAsO (*R* = La and Sm) was found to induce superconductivity [6-10]. Very recently, 5*d* metal Ir-doped *R*FeAsO (*R* = La and Sm) were also found to exhibit superconductivity [11-12]. In this letter, we report a bulk superconductivity in the 4*d* metal Rh-doped NdFeAsO polycrystalline samples with a maximum $T_C$ = 13 K. Based on the electrical resistivity and magnetization measurements, we construct a temperature- composition phase diagram for the Nd(Fe, Rh)AsO system.

Polycrystalline samples with a nominal composition of $NdFe_{1-x}Rh_xAsO$ (*x* = 0, 0.02, 0.04, 0.05, 0.08, 0.10, 0.12, 0.15, 0.20, 0.23, and 0.25) were synthesized by a solid state reaction method using NdAs and the mixtures of $Fe_{1-x}Rh_xO$ as starting materials. The mixture of Nd/As grains and that of Fe/Rh/$Fe_2O_3$ powders were mixed and ground, then sealed in the quartz tubes under vacuum. The mixture of NdAs was slowly heat to 500 $^oC$ for 10 h and then 900 $^oC$ for another 5 h, while the mixtures of $Fe_{1-x}Rh_xO$ were slowly heat to 500 $^oC$ for 10 h and then 800 $^oC$ for another 5 h. Stoichiometric amounts of NdAs and $Fe_{1-x}Rh_xO$ powders were thoroughly ground and pressed into pellets and then sintered in an evacuated quartz tube at 1150 $^oC$ for 40 h, with one intermediate stage of grinding and pressing to ensure homogeneity of the samples. All the material preparation processes were carried out in a glove box under argon atmosphere.

Powder X-ray diffraction (XRD) using Cu $K_\alpha$ radiation at room temperature was used to identify the phase structure of the samples. Figure 1 shows the powder XRD patterns for the representative $NdFe_{1-x}Rh_xAsO$ samples. One can see that the main diffraction peaks can be well indexed with the tetragonal ZrCuSiAs-type structure (space group *P*4/*nmm*), indicating the samples are almost of single phase. In the samples with x ≥ 0.20, a few weak peaks are observed due to the existence of a trace amount of impurity phases. The lattice parameters of the *a*-axis and *c*-axis as a function of the Rh content *x* for the $NdFe_{1-x}Rh_xAsO$ system are plotted in the inset of Fig. 1. The *a*-axis increases and the *c*-axis shrinks slightly with increasing the Rh content *x*. For the sample with *x* = 0.25, the *a*-axis and *c*-axis lattice parameters increases and decreases by 1.13 % and 2.21 %, respectively. A shrinkage of the *c*-axis in the *M*-doped *R*FeAsO (*R* = La and Sm) with $M$ = Co, Ni, and Ir [6-12], and an increase of the $a$-axis in the Ni and Ir-doped $R$FeAsO ($R$ = La and Sm) [10-12] were also observed.

We have measured the electrical resistivity as a function of temperature for the samples using the standard four-probe method (PPMS, Quantum Design). Figure 2 shows the temperature dependence of the resistivity, $\rho$ ($T$), normalized to $\rho$ (300K) for $NdFe_{1-x}Rh_xAsO$ samples. The room-temperature resistivity is found to decrease with increasing the Rh-doping level in the samples, similar to the observation in the Co-doped $R$FeAsO ($R$ = La and Sm) [6-7]. For the undoped NdFeAsO sample, the resistivity shows a fast drop at 147 K, which corresponds to the structural phase transition temperature $T_S$ (from tetragonal to orthorhombic), followed by an antiferromagnetic (AFM) transition [13]. For the samples with $x$ = 0.02 and 0.04, an anomaly in the $\rho$ ($T$) curves, which can be ascribed to the structural and magnetic phase transition at $T_S$, is determined to be at 120 and 90 K，respectively. For the samples with $0.05 \le x \le 0.25$, it is evident from Fig. 2 that no anomaly can be observed in the $\rho$ ($T$) curves, and superconducting transition occurs. Figure 3 (a) shows the enlarged $\rho$ ($T$) around $T_C$ for the representative samples. The maximum SC transition temperature $T_C$ = 13.6 K (the onset transition) or 12.8 K (the midpoint of the resistive transition) with a transition width $\Delta T_C$ = 0.8 K is observed in the $x$ = 0.10 sample. To confirm the bulk nature of the superconductivity in the samples, the magnetization as a function of temperature was measured with a vibrating sample magnetometer (PPMS, Quantum Design). Figure 3 (b) shows temperature dependence of the magnetization measured under condition of zero field cooling (ZFC) at $H$ = 20 Oe for the representative samples. The samples show strong diamagnetic signals, indicating bulk superconductivity of the Rh-doped samples. Therefore, it can be concluded that Rh doping in NdFeAsO suppresses the structural and magnetic phase transition, and induces the superconducting transition. Moreover, we notice that there exists a minimum in the $\rho$ ($T$) curves at $T_{min}$ for the SC samples in the normal state and also for the non-SC samples (see Fig. 2). It turns out that the electrical resistivty implies a metallic behavior for $T > T_{min}$ and a semiconductor-like behavior for $T < T_{min}$. The value for $T_{min}$ decreases monotonously with increasing the Rh content $x$. At present, it is unclear yet why there exists a minimum in the $\rho$ ($T$) curves. A similar phenomenon was also observed in the Co-doped $R$FeAsO ($R$ = La and Sm) system [7] and the Ir-doped SmFeAsO system [12].

Figure 3(c)-(e) show the magnetoresistive SC transition under different applied fields for the samples with $x = 0.10$, 0.15, and 0.20, respectively. One can see that, with increasing the applied magnetic fields, the SC transition is shifted towards lower temperatures, and the transition width is broadened considerably. Figure 3(f) shows the upper critical field $H_{c2}(T)$ as a function of temperature obtained from a determination of the midpoint of the resistive transition. Furthermore, we can estimate the value for the upper critical field $H_{c2}(0)$ using the Werthamer–Helfand–Hohenberg (WHH) formula, $H_{c2}(0) = -0.69T_C (dH_{c2} / dT)|_{Tc}$ [14]. For example, the slope $\mu_0(dH_{c2} / dT)|_{Tc}$ is determined to be $-2.94$ T/K for the $x = 0.10$ sample, and the value for $\mu_0 H_{c2}(0)$ is thus estimated to be about 26 T.

Figure 4 shows the temperature-composition phase diagram for the Nd(Fe, Rh)AsO system, which is constructed based on the transport and magnetization measurement data. The SC transition temperature $T_C$ is determined from the midpoint of the resistive transition. The data of $T_{min}$ are also plotted in this figure. The most striking feature in the Nd(Fe, Rh)AsO system is that no coexistence of antiferromagnetisim and superconductivity in an underdoped region is found, and a first-order-like transition from the AFM state to the SC state at a critical Rh-doping point $x_C \approx 0.045$ is observed. Our phase diagram for the $NdFe_{1-x}Rh_xAsO$ system is similar to that for the F-doped *R*FeAsO system with *R* = La, Pr, and Sm [15-18]. The general feature of the phase diagram for the *R*FeAsO-type superconductors is similar to that for the $AFe_2As_2$-type (*A* = Ca, Sr, and Ba) superconductors, but there is a difference in the underdoped region between the two types of superconductors. For the K-doped $AFe_2As_2$ (*A* = Sr and Ba) [19-21] and the *M*-doped $AFe_2As_2$ (*A* = Sr and Ba, *M* = Co, Ni, Rh, Pd) [22-28], a coexistence of antiferromagnetisim and superconductivity was observed in an underdoped region of the phase diagram. On the other hand, the highest $T_C = 17$ K observed in the *M*-doped *R*FeAsO system [6-12] is smaller than that in the *M*-doped $AFe_2As_2$ system ($T_C = 28$ K) [22-29], whereas the highest $T_C = 55$ K in F-doped *R*FeAsO system [1-5] is larger than that in K-doped $AFe_2As_2$ system ($T_C = 38$ K) [19-21, 30]. Therefore, further experimental and theoretical work is needed for the interpretation of the above-mentioned different properties between the *R*FeAsO-type superconductors and the $AFe_2As_2$-type superconductors.

In summary, we have shown that Rh doping in NdFeAsO suppresses the structural and magnetic phase transition, and induces superconductivity with a maximum $T_C = 13$ K. A temperature-composition phase diagram has been established for the $NdFe_{1-x}Rh_xAsO$ system based on the electrical resistivity and magnetization measurements. A first-order-like transition from the AFM state to the SC state at a critical Rh-doping point $x_C \approx 0.045$ appears in the phase diagram. The value for $\mu_0 H_{c2}(0)$ is estimated to be 26 T for the $x = 0.10$ sample.

**Acknowledgements** The authors thank W. Yu for helpful discussions. This work was supported by the National Basic Research Program of China (Contract No. 2007CB925001) and by the NSFC (Grant Nos. 10874244 and 10974254).

**Figure captions**

**Figure 1** (a) Powder x-ray diffraction patterns for the representative $NdFe_{1-x}Rh_xAsO$ samples. (b) Lattice parameters *a* and *c* as a function of Rh content *x*.

**Figure 2** Temperature dependence of resistivity $\rho$ ($T$), normalized to $\rho$ (300 K), for $NdFe_{1-x}Rh_xAsO$. The curves are vertically offset and separated by 0.2 units for clarity. Arrows indicate the position of the structural and magnetic phase transition $T_S$.

**Figure 3** (a) Enlarged temperature dependence of resistivity around $T_C$, and (b) Temperature dependence of magnetization for the representative $NdFe_{1-x}Rh_xAsO$ samples. The data was measured under condition of zero field cooling (ZFC) at $H$ = 20 Oe. (c)-(e) Temperature dependence of the resistivity, $\rho$ ($T$), under different applied fields from 0 to 10 T for (c) $x = 0.10$, (d) $x = 0.15$, and (e) $x = 0.20$ samples. (f) The upper critical filed $H_{c2}$ ($T$) as a function of temperature.

**Figure 4** Temperature-composition phase diagram for the $NdFe_{1-x}Rh_xAsO$ system.

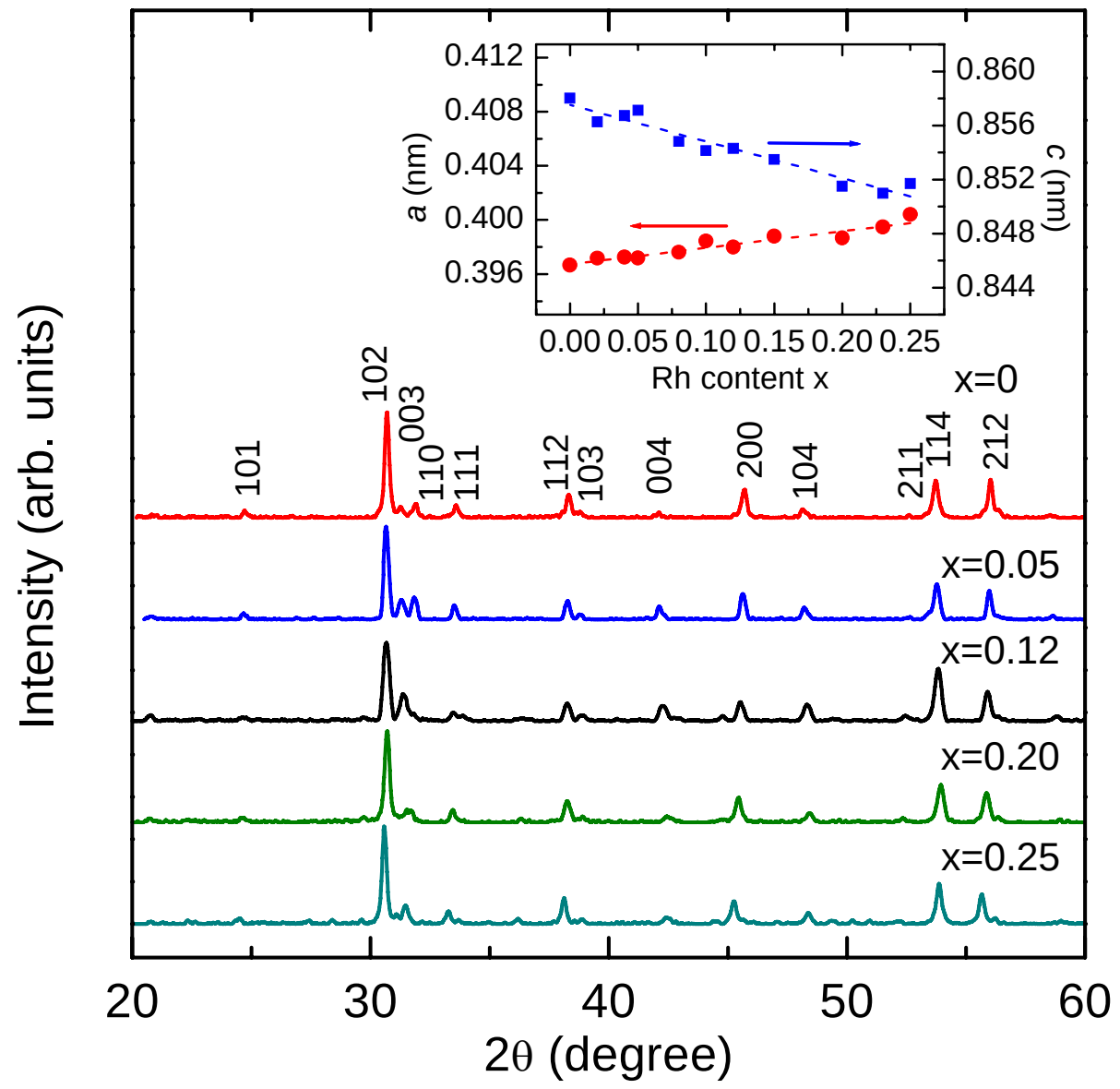

Figure 1

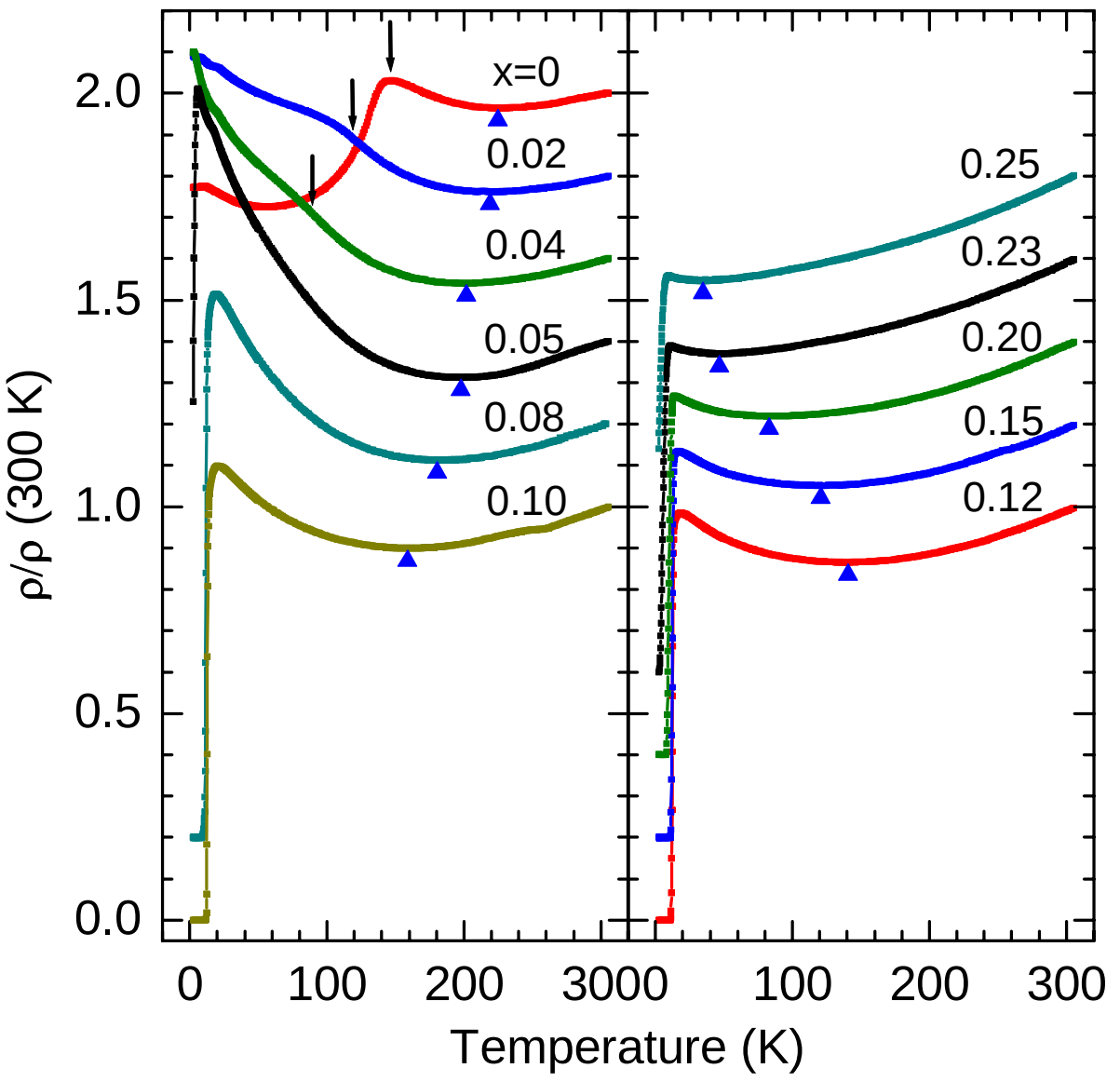

Figure 2

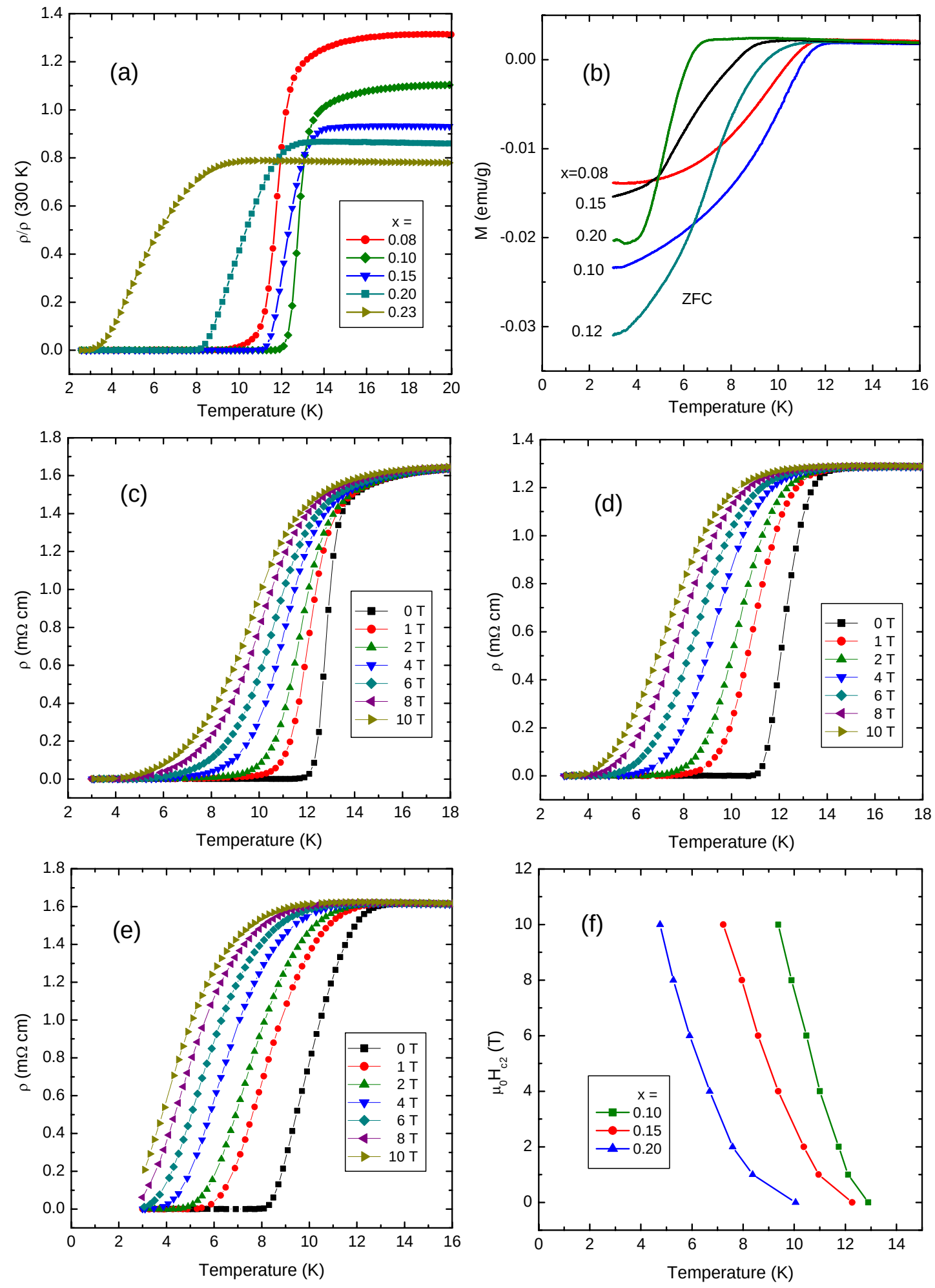

Figure 3

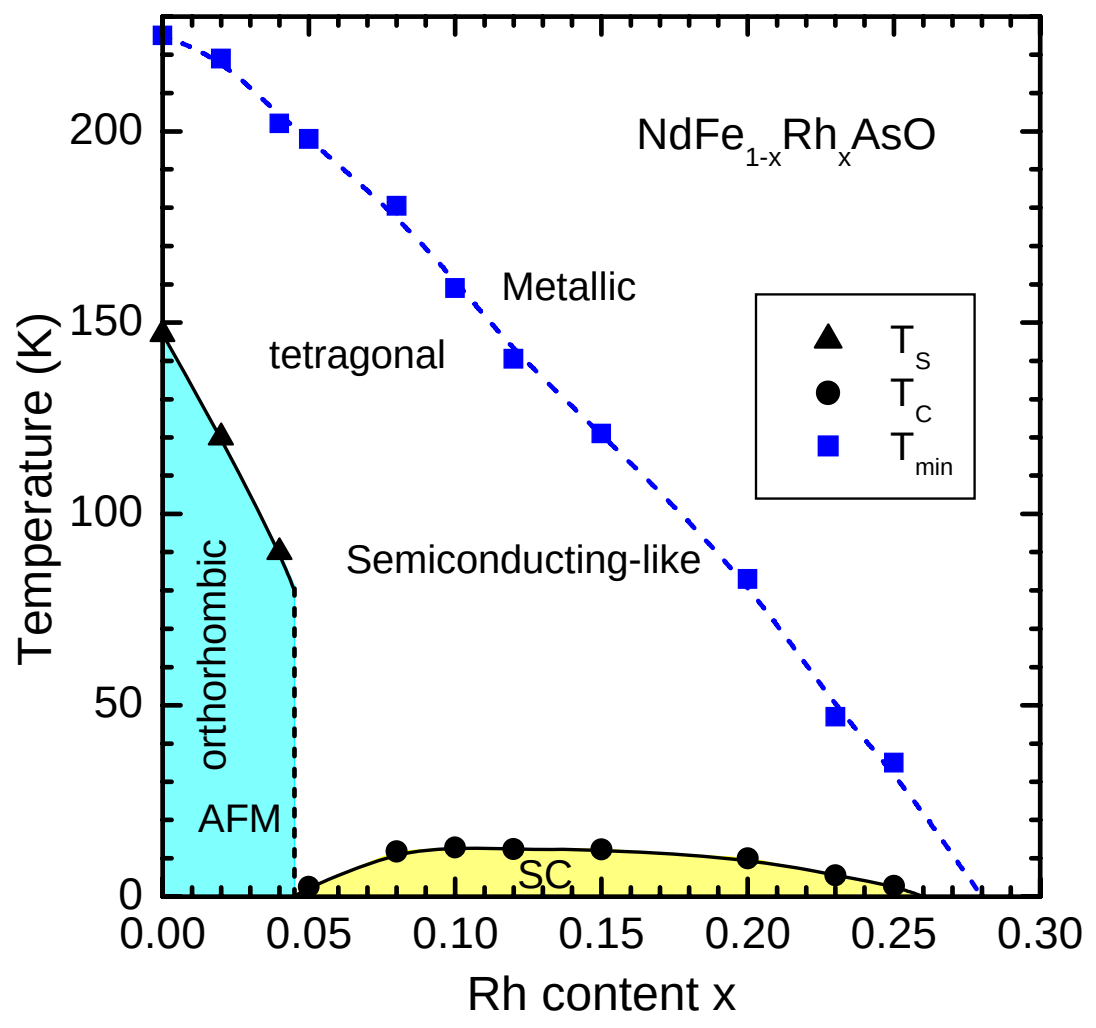

Figure 4